\documentclass[12pt,twocolumns]{iopart}
\usepackage{graphicx}
\begin{document}
\title{Atom-atom entanglement generated at early times by two photon emission}
\author{Juan Le\'on and Carlos Sab\'in}
\address{Instituto de F\'isica Fundamental, CSIC, Serrano 113-bis, 28006 Madrid, Spain}
\eads{\mailto{leon@imaff.cfmac.csic.es} and \mailto{csl@imaff.cfmac.csic.es}}
\begin{abstract}
We analyze entanglement generation between a pair of neutral two level atoms that are initially excited in a common
electromagnetic vacuum. The nonlocal correlations that appear due to the interaction with the field can become entanglement when
the field state is known. We distinguish two different situations: in the first, the field remains in the vacuum state, and
second two photons are present in the final state. In both cases we study the dependence of the entanglement with time and
interatomic distance, at ranges related with locality issues.
\end{abstract}

\ \ \\[3mm]

As is well known, an atom $B$ cannot become excited by a physical signal coming from other atom $A$ placed at a distance $r$ at
a time $t< t_r=r/c$. This does not prevent the emergence of correlations between these two atoms at $t<t_r$. In particular, when
the final state of the field (and/or atom $A$) is fixed, atom $B$ can become excited even when both atoms were initially (at
$t=0$) uncorrelated and the time elapsed was smaller than $t_r$. The specification of this final state is highly nonlocal and a
finite excitation probability $P_B(t)$ for atom $B$ at times $t$ earlier than $t_r$ does not imply causality violation but is an
artifact of that nonlocality. As shown in \cite{powerthiru}, causality is fully restored after summing over all the accessible
final states for atom $A$ and the electromagnetic field. Now only the state of $B$ is accounted for; this is a local property,
no wonder that $P_B(t)=0$ for $t<t_r$ in this case. In the language of Quantum Information, the system is nonsignaling
\cite{gisin}. It is well known that the mere existence of correlations is not enough in order to send information between the
parties: classical communication is always needed to transform entanglement into information.

Due to the existence of these nonlocal correlations, the reduced state of the atoms at a time $t$ is a mixed classically
correlated state, even for $c\, t<r$ \cite{conjuan}. The generation of a correlated state from an uncorrelated one only by local
interactions could be blamed on the fact that the field vacuum is an entangled state between spacelike separated regions
\cite{summerswerner,summerswernerII,reznik} or on the finiteness of the Feynman propagator for $c\, t<r$ \cite{franson}. There
are at least two ways to obtain entanglement from these correlations: introducing a suitable time dependent coupling between the
field and the atoms (like in \cite{reznik,reznikII} with a scalar field and two classical detectors) or fixing the field state
\cite{conjuan, conjuanII, conjuanIII}. Here we will show new examples of the latter case: we will start with both atoms excited
in the electromagnetic vacuum, and after a interaction time $t$ we will consider the following field states:
\begin{itemize} \item[i)] no photons are present,\item[ii)]  two photons are detected. \end{itemize}

We shall study the dependence of the entanglement with $t$ and $r$, both for $r> c\,t$ and $r<c\,t$.

We assume that the wavelengths relevant in the interaction with the atoms, and the separation between them, are much longer than
the atomic dimensions. The dipole approximation, appropriate to these conditions,  permits the splitting of the system
Hamiltonian into two parts $H = H_0 + H_I$ that are separately gauge invariant. The first part is the Hamiltonian in the absence
of interactions other than the potentials that keep $A$ and $B$ stable, $H_0 = H_A + H_B + H_{\mbox{field}}$. The second
contains all  the interaction of the atoms with the field
\begin{equation}
H_I = - \frac{1}{\epsilon_0}\sum_{n=A,B} \mathbf{d}_n(\mathbf{x}_n,t)\,\mathbf{D}(\mathbf{x}_n,t) \label{a},
\end{equation}
where $\mathbf{D}$ is the electric displacement field, and $\mathbf{d}_n \,=\,\sum_i\, e\,\int d^3 \mathbf{x}_i\,
\langle\,E\,|\,(\mathbf{x}_i-\mathbf{x}_n)\,|\,G\,\rangle$ is the electric dipole moment of atom $n$, that we will take here as
real and of equal magnitude for both atoms $(\mathbf{d}=\mathbf{d_A}=\mathbf{d_B})$, $|\,E\,\rangle$ and $|\,G\,\rangle$ being
the excited and ground states of the atoms, respectively.

In what follows we choose a system  given initially by the product state, $|\,\psi\,\rangle_0\,=\,
|\,E\,E\,\rangle\cdot|\,0\,\rangle$ in which atoms $A$ and $B$ are in the excited state $|\,E\,\rangle$ and the field in the
vacuum state $|\,0\,\rangle$. The system then evolves under the effect of the interaction during a lapse of time $t$ into a
state:
\begin{equation}
|\,\psi\,\rangle_t = e^{-i\, \int_0^t\,dt'\, H_I\,(t')/\hbar}\,|\,\psi\,\rangle_0 \label{b}
\end{equation}
that,  to  order $\alpha$, can be given in the interaction picture as
\begin{eqnarray}
|\mbox{atom}_1,\mbox{atom}_2,\mbox{field}\rangle_{t} =  ((1+a)\,|\,E\,E\rangle + b\,|\,G\,G\rangle)\,|\,0\rangle\nonumber\\
 +(u_A\,|\,G\,E\,\rangle+ u_B\,|\,E\,G\,\rangle)\,|\,1\,\rangle+
(f\,|\,E\,E\rangle+ g\,|\,G\,G\rangle)\,|\,2\rangle\  \label{c}
\end{eqnarray}
where
\begin{eqnarray}
a&=&\frac{1}{2}\langle0|T(\mathcal{S}_A^+ \mathcal{S}_A^- + \mathcal{S}_B^+\mathcal{S}_B^-)|0\rangle,\, b=
\langle0|T(\mathcal{S}^-_B\mathcal{S}^-_A)|0\rangle\nonumber\\
u_A\,&=&\,\langle\,1\,|\, \mathcal{S}^-_A\,|\,0\,\rangle,\, u_B\,=\,\langle\,1\,|\,
\mathcal{S}^-_B\,|\,0\,\rangle \label{d}\\
f&=&\frac{1}{2}\langle2|T(\mathcal{S}_A^+ \mathcal{S}_A^- +\mathcal{S}_B^+\mathcal{S}_B^-)|0\rangle,\,
g=\langle2|T(\mathcal{S}^-_B \mathcal{S}^-_A)|0\rangle,\nonumber
\end{eqnarray}
being $\mathcal{S}\,=\,- \frac{i}{\hbar}  \int_0^t\, dt'\, H_{I}(t')=\mathcal{S}^{+}\, +\, \mathcal{S}^{-}$, $T$ being the time
ordering operator and $|\,n\,\rangle,\,\, n=\,0,\,1,\,2$ is a shorthand for the state of $n$ photons with definite momenta and
polarizations, i.e. $|\,1\,\rangle\,=\,|\mathbf{k},\, \mathbf{\epsilon}_{\lambda}\,\rangle$, etc. The superscript signs in
$\mathcal{S}^{\pm}$, to be defined in \ref{e} below, are associated to the energy difference between the initial and final
atomic states of each transition. Among all the terms that contribute to the final state (\ref{c}) only $b$ corresponds to
interaction between both atoms. This would change at higher orders in $\alpha$. Here, $a$ describes intra-atomic radiative
corrections, $u_A$ and $u_B$ single photon emission by one atom, and $g$ by both atoms, while $f$ corresponds to two photon
emission by a single atom. In Quantum Optics, virtual terms like $b$ and $f$ which do not conserve energy and appear only at
very short times, are usually neglected by the introduction of a rotating wave approximation (RWA). But here we are interested
in the short time behavior, and therefore all the terms must be included, as in \cite{powerthiru,milonni,compagnoI}. Actually,
only when all these virtual effects are considered, it can be said properly that the probability of excitation of atom $B$ is
completely independent of atom $A$ when $r>c\,t$ \cite{milonni, compagnoI}.

Finally, in the dipole approximation the actions $\hbar\, \mathcal{S}^{\pm}$  in (\ref{d}) reduce to
\begin{eqnarray}
\mathcal{S}^{\pm}\,=\,- \frac{i}{\hbar}  \int_0^t\, dt' \: e^{\pm i\Omega t'}\, \mathbf{d}\,\mathbf{E}(\mathbf{x},t')\label{e}
\end{eqnarray}
where  $\Omega = \omega_E -\omega_G$ is the transition frequency, and we are neglecting atomic recoil.  (\ref{e}) depends on the
atomic properties $\Omega$ and $\mathbf{d}$, and on the interaction time $t$. In our calculations we will take $(\Omega
|\mathbf{d}|/e c) = 5\,\cdot 10^{-3}$, which is of the same order as the 1s $\rightarrow$ 2p transition in the hydrogen atom,
consider $\Omega\,t > 1$, and analyze the cases $(r/c\,t)\simeq 1$ near the time $t=t_r$ where one atom could begin to receive
signals from the other.

Given a definite field state $|\,n\,\rangle$ the pair of atoms is in a  projected pure two qubits state as shown in (\ref{c}).
We will denote these states by $|\,A,B,n\,\rangle$, $\rho_{AB}^{(n)}\,=\,|A,B,n\rangle\,\langle A,B,n\,|$ and will compute the
concurrence $\mathcal{C}^{(n)}$ \cite{wootters}:
\begin{equation}
\mathcal{C}^{(n)}= max\{0, \sqrt{\lambda_i}-\sum_{j\neq i}\, \sqrt{\lambda_j}\} \label{g}
\end{equation}
where $\lambda_i$ the largest of the eigenvalues $\lambda_j$ ($j=1,...,4$) of
$[(\sigma^A_y\otimes\sigma^B_y)\rho_{AB}^{(n)*}(\sigma^A_y\otimes\sigma^B_y)]\rho_{AB}^{(n)}$.

We begin with the case $n=0$, where the field is in the vacuum state and, following (\ref{c}), the atoms are in the projected
pure state $|\,A\,B\,0\,\rangle=((1\,+\,a)\,|\,E\,E\,\rangle + b\,|\,G\,G\,\rangle) / c_0$, where $c_0\,=\, \sqrt{|1\,+\,a|^2
\,+\, |b|^2}$ is the normalization, giving a concurrence
\begin{equation}
 \mathcal{C}^{(0)} \,=\, 2\,|b|\,|\,1\,+\,a\,|/c_0^2\ .\label{h}
\end{equation}
The computation of $a$ and $b$ can be performed following the lines given in \cite{conjuan}, where they were computed for the
case of a initial atomic state $|\,E\,G\,\rangle$. We will consider that the dipoles are parallel along the $z$ axis, while the
atoms remain along the $y$ axis. Under that conditions, using the dimensionless variables $x=r/c\,t$ and $z=\Omega\,r/c$:
\begin{eqnarray}
a&=&\frac{4\,i\,K\,z^3}{3\,x}\,(\ln{|1-\frac{z_{max}}{z}|}\,+\,2\,i\,\pi),\nonumber\\
b&=&\frac{\alpha\,d_i\,d_j}{\pi\,e^2}(-\mathbf{\nabla}^2\delta_{ij}+\nabla_i\nabla_j)\,I, \label{i}
\end{eqnarray}
with $K=\alpha\,|\,\mathbf{d}\,|^2/(e^2\,r^2)$ and $I=I_+\,+\,I_-$, where:
\begin{eqnarray}
I_{\pm}&=&\frac{-i\,e^{-i\frac{z}{x}}}{2\,z}\,[\,\pm\,2\cos(\,\frac{z}{x}\,)\,e^{\pm\,i\,z}\,Ei(\mp\,i\,z)
+\,e^{-i\,z\,(1\pm\frac{1}{x})}\nonumber\\&\ &
Ei(i\,z\,(1\pm\frac{1}{x}))\,-\,e^{i\,z\,(1\pm\frac{1}{x})}\,Ei(-i\,z\,(1\pm\frac{1}{x}))\,]\label{j}
\end{eqnarray}
for $x>1$ with the additional term $-2\,\pi\,i\,e^{i\,z\,(1-1/x)}$ for $x<1$. We use the conventions and tables of
\cite{bateman}.

We show in Fig. 1 the concurrence $\mathcal{C}^{(0)}$  (\ref{h}) for $x$ near 1 for given values of $z$. Like the case where
$|\,E\,G\,\rangle$ is the initial atomic state, $\mathcal{C}^{(0)}$ jumps at $x=1$ and has different behaviors at both sides.
\begin{figure}[h]
\includegraphics[width=0.5\textwidth]{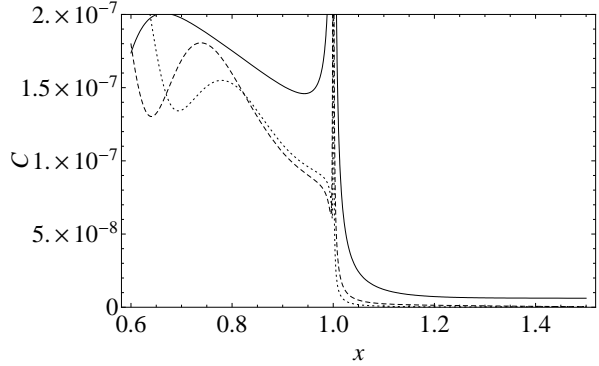}
\caption{Concurrence $\mathcal{C}^{(0)}$ of the atomic state in the e.m. vacuum  $\rho_{AB}^{(0)}$ as a function of $x=(r/c\,t)$
for $z=(\Omega r/c)=$ 5 (solid line), 10 (dashed line) and 15 (dotted line). The height of the peak is $\mathcal{C}^{(0)}=1$.
$x\rightarrow0$ ($t\rightarrow\infty$) is the region usually considered in Quantum Optics.}
\end{figure}

In Fig. 2 the concurrence is sketched as a function of $z$ for given values of $\Omega\,t=z/x$. The tiny values of the
concurrence for the region $z>\Omega\,t$ (which corresponds to $x>1$), diminish as $t$ grows and will eventually vanish, since
$b$ is a non-RWA term.
\begin{figure}[h]
\includegraphics[width=0.5\textwidth]{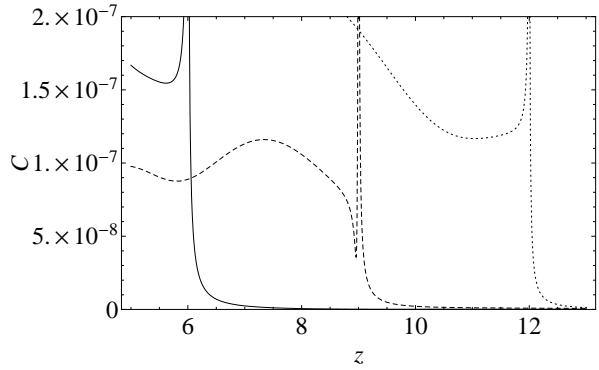}
\caption{Concurrence $\mathcal{C}^{(0)}$ of the atomic state in the e.m. vacuum  $\rho_{AB}^{(0)}$ as a function of
$z=(\Omega\,r/c)$ for $z/x=\Omega\,t=$ 6 (solid line), 9 (dashed line) and 12 (dotted line). $x>1$ amounts to $z>\Omega\,t$ in
each case. }
\end{figure}

The case $n=1$ analyzed in \cite{conjuanII}, shows atom-atom correlations due to the uncertainty of the photon source. Here, we
will focus on the two photon case. The final atomic state $|\,A\,B\,2\rangle=(f\,|\,E\,G\,\rangle\,+\,g\,|\,G\,E\,\rangle)/c_2$,
with $c_2\,=\,\sqrt{|\,f\,|^2\,+\,|\,g\,|^2}$, is in the same subspace as for $n\,=\,0$. The normalization $c_2$ is
$\mathcal{O}(\alpha)$ like the expectation values $f$, $g$, so that all the coefficients in $\rho^{(2)}$ may be large.
Therefore, although the probability of attaining this state is small, the correlations are not. The concurrence is
\begin{equation}
\mathcal{C}^{(2)}=2 |\,f\,g^*\,|/c_2^2.\label{k}
\end{equation}
We find that:
\begin{eqnarray}
f&=&\theta(t_1-t_2)(\,v_A\,(t_1)\,u'_A\,(t_2)\,+\,u_A\,(t_1)\,v'_A\,(t_2)\nonumber\\
&+&\,v_B\,(t_1)\,u'_B\,(t_2)\,+\,u_B\,(t_1)\,v'_B\,(t_2)\,)\label{l}\\
g&=&u_B\,u'_A\,+\,u_A\,u'_B\nonumber
\end{eqnarray}
with $v_A\,=\,\langle\,1\,|\, \mathcal{S}^+_A\,|\,0\,\rangle$ and $v_B\,=\,\langle\,1\,|\, \mathcal{S}^+_B\,|\,0\,\rangle$. The
primes account for the two single photons, i.e.
$|\,2\,\rangle=|\mathbf{k}\,\mathbf{\epsilon}_{\lambda},\,\mathbf{k'}\,\mathbf{\epsilon}_{\lambda'}\,\rangle$. The quantities
$|\,u_A\,|^2\,=\,|\,u_B\,|^2\,=\,|\,u\,|^2$, $|\,v_A\,|^2\,=\,|\,v_B\,|^2\,=\,|\,v\,|^2$, $l\,=\,u_A\,v_B^*\,=\,u_B\,v_A^*$,
$u\,v^*\,=\,u_A^*\,v_A^*=\,u_B\,v_B^*$, $u_B\,u_A^*$ and $v_A\,v_B^*$ have been computed in \cite{conjuan}.

In Fig. 3 we show $\mathcal{C}^{(2)}$ in front of $x$ for given values of $z$. When $x\rightarrow0$ ($t\rightarrow\infty$, i.e.
the Quantum Optics regime), $f$ vanishes and the final atomic state would be the separable state $|\,G\,G\,\rangle$, with zero
concurrence. Entanglement is sizeable for $x>1$, and could be maximized if a particular two photon state was detected
\cite{conjuanIII}.
\begin{figure}[h]
\includegraphics[width=0.5\textwidth]{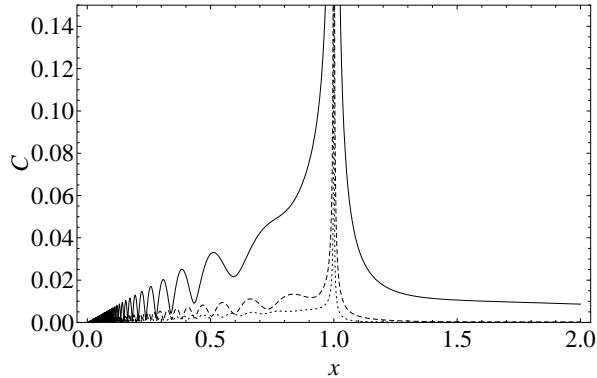}
\caption{Concurrence $\mathcal{C}^{(2)}$ of the atomic state with $n=2$ photons $\rho_{AB}^{(2)}$ in front of $x=r/(c\,t)$ for
$z=\Omega\,r/c=$ 5 (solid line), 10 (dashed line) and 15 (dotted line).}
\end{figure}

In Fig. 4,  $\mathcal{C}^{(2)}$ is sketched as a function of $z$ for given values of $\Omega\,t=z/x$. Again, the concurrence for
the region $z>\Omega\,t$ ($x>1$), diminish as $t$ grows and will eventually vanish, since it is due to $f$, which is a non-RWA
term.
\begin{figure}[h]
\includegraphics[width=0.5\textwidth]{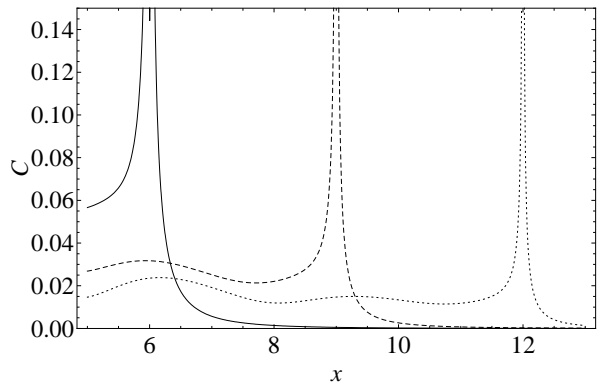}
\caption{Concurrence $\mathcal{C}^{(2)}$ of the atomic state with $n=2$ photons $\rho_{AB}^{(2)}$ in front of $z=\Omega\, r/c$
for $\Omega\,t=z/x=$ 6 (solid line), 9 (dashed line) and 12 (dotted line).}
\end{figure}
Interestingly, as we noted in \cite{conjuanII} for the single photon emission, $x=1$ is a singular point that divides the
spacetime into two different regions. This occurs, as in \cite{conjuanII}, even if in these cases $t$ is not the propagation
time of any physical signal between the atoms. This effect comes from the appearance of effective interaction terms like $l$,
that would be missing if we could discriminate the source of emission of each photon.

As a summary, we have computed the concurrence generated between a pair of neutral two level atoms by the interaction with the
electromagnetic field, in order to study the role of locality in the growing up of quantum correlations. We have considered the
initial state with both atoms excited and the field in the vacuum, and two different situations after a time $t$: the field
remains in the vacuum and two photons have been emitted. In the former case, entanglement is generated by the interaction
amplitude $b$. For $r<c\,t$ photon exchange make correlations grow during a short time, before vanish eventually due to the
non-RWA nature of $b$. For $r>c\,t$ although $b$ is nonzero its amplitude is negligible compared with $r<c\,t$. The situation is
different when two photons are present. There is a relevant probability that both photons come from the same atom, which in our
scheme is represented by the non-RWA amplitude $g$. As long as we cannot distinguish the source of emission of each photon,
concurrence is sizeable for $r>c\,t$ and $r<c\,t$, and will eventually vanish. Although there are no interaction terms between
the atoms, the indistinguishability construes effective interaction terms and $r=c\,t$ plays the role of a frontier between two
different spacetime regions.

\ack
This work was supported by Spanish MEC FIS2005-05304 and CSIC 2004 5 OE 271 projects.

\end{document}